\documentclass[aps,prl,twocolumn,showpacs,psfig,superscriptaddress,longbibliography]{revtex4-2}

\usepackage{textcomp}
\usepackage{times}
\usepackage{graphicx}
\usepackage{float}
\usepackage{latexsym,amsmath,amssymb,bm,euscript}
\usepackage{color}
\usepackage{subfigure}
\usepackage{epstopdf}
\usepackage[colorlinks=true,linkcolor=blue,citecolor=blue]{hyperref}
\usepackage{hyperref}
\usepackage{soul}
\usepackage[normalem]{ulem}
\usepackage{mathrsfs}
\usepackage{amsmath}
\usepackage{lettrine}
\usepackage{xspace}
\usepackage{textcomp}

\begin{document}
\title{Spin-supersolidity induced quantum criticality and magnetocaloric effect in the triangular-lattice antiferromagnet Rb$_2$Co(SeO$_3$)$_2$}

\author{Yi Cui}
\thanks{These authors contributed equally to this study.}
\affiliation{School of Physics and Beijing Key Laboratory of
Opto-electronic Functional Materials $\&$ Micro-nano Devices, Renmin
University of China, Beijing, 100872, China}

\author{Zhanlong Wu}
\thanks{These authors contributed equally to this study.}
\affiliation{School of Physics and Beijing Key Laboratory of
Opto-electronic Functional Materials $\&$ Micro-nano Devices, Renmin
University of China, Beijing, 100872, China}

\author{Zhongcen Sun}
\thanks{These authors contributed equally to this study.}
\affiliation{School of Physics and Beijing Key Laboratory of
Opto-electronic Functional Materials $\&$ Micro-nano Devices, Renmin
University of China, Beijing, 100872, China}

\author{Kefan Du}
\affiliation{School of Physics and Beijing Key Laboratory of
Opto-electronic Functional Materials $\&$ Micro-nano Devices, Renmin
University of China, Beijing, 100872, China}

\author{Jun Luo}
\affiliation{Institute of Physics, Chinese Academy of Sciences, and Beijing National Laboratory for Condensed Matter Physics, 100190, Beijing, China}

\author{Shuo Li}
\affiliation{Institute of Physics, Chinese Academy of Sciences, and Beijing National Laboratory for Condensed Matter Physics, 100190, Beijing, China}

\author{Jie Yang}
\affiliation{Institute of Physics, Chinese Academy of Sciences, and Beijing National Laboratory for Condensed Matter Physics, 100190, Beijing, China}

\author{Jinchen Wang}
\email{jcwang\_phys@ruc.edu.cn}
\affiliation{School of Physics and Beijing Key Laboratory of
Opto-electronic Functional Materials $\&$ Micro-nano Devices, Renmin
University of China, Beijing, 100872, China}

\author{Rui Zhou}
\email{rzhou@iphy.ac.cn}
\affiliation{Institute of Physics, Chinese Academy of Sciences, and Beijing National Laboratory for Condensed Matter Physics, 100190, Beijing, China}

\author{Qian Chen}
\affiliation{Institute for Solid State Physics, The University of Tokyo, Kashiwa, Chiba, 277-8581, Japan}

\author{Yoshimitsu Kohama}
\affiliation{Institute for Solid State Physics, The University of Tokyo, Kashiwa, Chiba, 277-8581, Japan}

\author{Atsuhiko Miyata}
\email{a-miyata@issp.u-tokyo.ac.jp}
\affiliation{Institute for Solid State Physics, The University of Tokyo, Kashiwa, Chiba, 277-8581, Japan}

\author{Zhuo Yang}
\email{zhuo.yang@issp.u-tokyo.ac.jp}
\affiliation{Institute for Solid State Physics, The University of Tokyo, Kashiwa, Chiba, 277-8581, Japan}

\author{Rong Yu}
\affiliation{School of Physics and Beijing Key Laboratory of
Opto-electronic Functional Materials $\&$ Micro-nano Devices, Renmin
University of China, Beijing, 100872, China}
\affiliation{Key Laboratory of Quantum State Construction and
Manipulation (Ministry of Education), Renmin University
of China, Beijing, 100872, China}

\author{Weiqiang Yu}
\email{wqyu\_phy@ruc.edu.cn}
\affiliation{School of Physics and Beijing Key Laboratory of
Opto-electronic Functional Materials $\&$ Micro-nano Devices, Renmin
University of China, Beijing, 100872, China}
\affiliation{Key Laboratory of Quantum State Construction and
Manipulation (Ministry of Education), Renmin University
of China, Beijing, 100872, China}


\begin{abstract}

We performed high-field magnetization, magnetocaloric effect (MCE), and NMR measurements on the Ising triangular-lattice antiferromagnet Rb$_2$Co(SeO$_3$)$_2$. The observations of the 1/3-magnetization plateau, the split NMR lines, and the thermal activation behaviors of the spin-lattice relaxation rate $1/T_1$
between 2~T and 15.8~T provide unambiguous evidence of a gapped up-up-down (UUD) magnetic ordered phase.
For fields between 15.8~T and 18.5~T, the anomaly in the magnetic susceptibility, the slow saturation of the NMR line spectral ratio
with temperature, and the power-law temperature dependence of $1/T_1$ suggest the ground state to be a spin supersolid with gapless spin excitations.
With further increasing the field, the Gr\"{u}neisen ratio, extracted from the MCE data, reveals
a continuous quantum phase transition at $H_{\rm C}\approx$ 19.5~T and a universal quantum critical scaling with the exponents ${\nu}z~\approx~$1.
Near $H_{\rm C}$, the large high-temperature MCE signal and the broad peaks in the NMR Knight shift and $1/T_1$, manifest the strong spin fluctuations
driven by both magnetic frustration and quantum criticality. These results
establish Rb$_2$Co(SeO$_3$)$_2$ as a candidate platform for  cryogenic magnetocaloric cooling.

\end{abstract}

\maketitle

{\bf Introduction.}
Frustrated quantum magnets provide a fertile platform for the emergence of exotic quantum phases~\cite{Balents_Nature_2010} and unconventional quantum criticality~\cite{Senthil_Science_2004}.
In particular, the spin-$1/2$ triangular-lattice antiferromagnet (TLAFM) has attracted significant interest due to the geometric frustration, which promotes rich phases and field-induced quantum phenomena~\cite{Yamamoto_PRL_2014}, such as quantum spin liquid~\cite{Balents_Nature_2010}, spin nematics~\cite{Tsunetsugu_JPSJ_2006,Sheng_NM_2025}, and spin supersolid (SS)~\cite{Melko_PRL_2005,Wessel_PRL_2005,Boninsegni_PRL_2005,Heidarian_PRL_2005,Wang_PRL_2009,Heidarian_PRL_2010,Yamamoto_PRL_2014,LiW_Nature_2024}.
Supersolidity refers to a quantum state with simultaneous presence of
a superfluid order that breaks the U(1) symmetry and a solid order
that breaks the translational symmetry.
Since originally proposed in the context of $^4$He
~\cite{Andreev_SPU_1969,Thouless_AP_1969,Leggett_PRL_1970}, this concept
has been extended to various systems including ultracold atomic gases~\cite{Li_Nature_2017,Leonard_Nature_2017,Tanzi_Nature_2019,Norcia_Nature_2021} and quantum spin systems~\cite{Melko_PRL_2005,Wessel_PRL_2005,Boninsegni_PRL_2005,Heidarian_PRL_2005,Wang_PRL_2009,Heidarian_PRL_2010,Yamamoto_PRL_2014,LiW_Nature_2024}. Although substantial theoretical works, particularly on hard-core bosons in triangular lattices~\cite{Melko_PRL_2005,Wessel_PRL_2005,Boninsegni_PRL_2005,Heidarian_PRL_2005,Wang_PRL_2009}, suggest routes to stabilize supersolids, their experimental realization, including in $^4$He, has been challenging~\cite{Kim_PRL_2012,Boninsegni_RMP_2012}, with a definitive demonstration reported only recently in a photonic system~\cite{Trypogeorgos_Nature_2025}.

As a natural extension in quantum magnets, the spin SS, referring to a spatially hybrid coplanar order with both transverse and longitudinal spin components, has been theoretically proposed for the spin-1/2 TLAFM with strong Ising anisotropy, which is described by an XXZ model~\cite{Jiang_PRB_2009,Heidarian_PRL_2010,Yamamoto_PRL_2014,Gao_npj_2022}. With a longitudinal field applied, the interplay of magnetic frustration and quantum fluctuations
gives rise to two spin SS phases, with the Y-type and the V-type magnetic structures, respectively, for fields just below and above the up-up-down (UUD) phase with 1/3 magnetization plateau~\cite{Yamamoto_PRL_2014,Sellmann_PRB_2015,Gao_npj_2022}.  Recently, gapless excitations associated with these spin SS phases were experimentally reported in the Ising TLAFM Na$_2$BaCo(PO$_4$)$_2$~\cite{Sheng_2022_PNAS,JinWT_PRM_2024,LiW_Nature_2024,Zhang_arxiv_2024,Xu_arxiv_2025}. In particular, the spin SS state is found to generate strong entropy fluctuations, which could give rise to a giant magnetocaloric effect (MCE)~\cite{LiW_Nature_2024}. It is well-known that competing orders in frustrated magnets can produce a substantial MCE over a broad temperature range. However,
for a SS, the coherence among its two ordered components and its impact to quantum criticality and spin fluctuations near the critical field remain elusive.  This naturally raises the question of whether enhancing competing fluctuations by tuning the SS toward a quantum critical point (QCP) can lead to highly efficient magnetocaloric cooling~\cite{Zhitomirsky_PRB_2003}.

More recently, another class of layered Ising TLAFM, $A$$_2$Co(SeO$_3$)$_2$ ($A$=K, Rb), have been synthesized~\cite{ZhongYD_PRM_2020}, and some of them
exhibit features of quantum spin liquid at low fields~\cite{ZhuM_PRL_2024, Mila_JCCMP_2024}.
Inelastic neutron scattering measurements on K$_2$Co(SeO$_3$)$_2$ revealed strong Ising exchange anisotropy, with dominant out-of-plane coupling $J_z~{\approx}$~3.1~meV and much weaker in-plane exchange $J_{xy}~{\approx}$~0.217~meV~\cite{ZhuM_PRL_2024,Zhu_npjQM_2024}.
The ground state of K$_2$Co(SeO$_3$)$_2$ at zero field is also proposed to host a Y-structure spin SS ~\cite{ZhuM_PRL_2024,ChenT_arxiv_2024}, and the spin excitations exhibit a number of intriguing features, including a low-energy continuum with a roton-like minimum at the $M$ point and a pseudo-Goldstone mode at the $K$ point,
which reflect strong quantum fluctuations and possible spin fractionalization~\cite{Zhu_npjQM_2024}. A 1/3 magnetization plateau is identified for
fields between 1 and 18~T, and a second putative spin SS phase between 18 and 21~T has also been suggested~\cite{ChenT_arxiv_2024}, though
remains to be verified spectroscopically.

\begin{figure}[t]
    \centering
    \includegraphics[width=8.5cm]{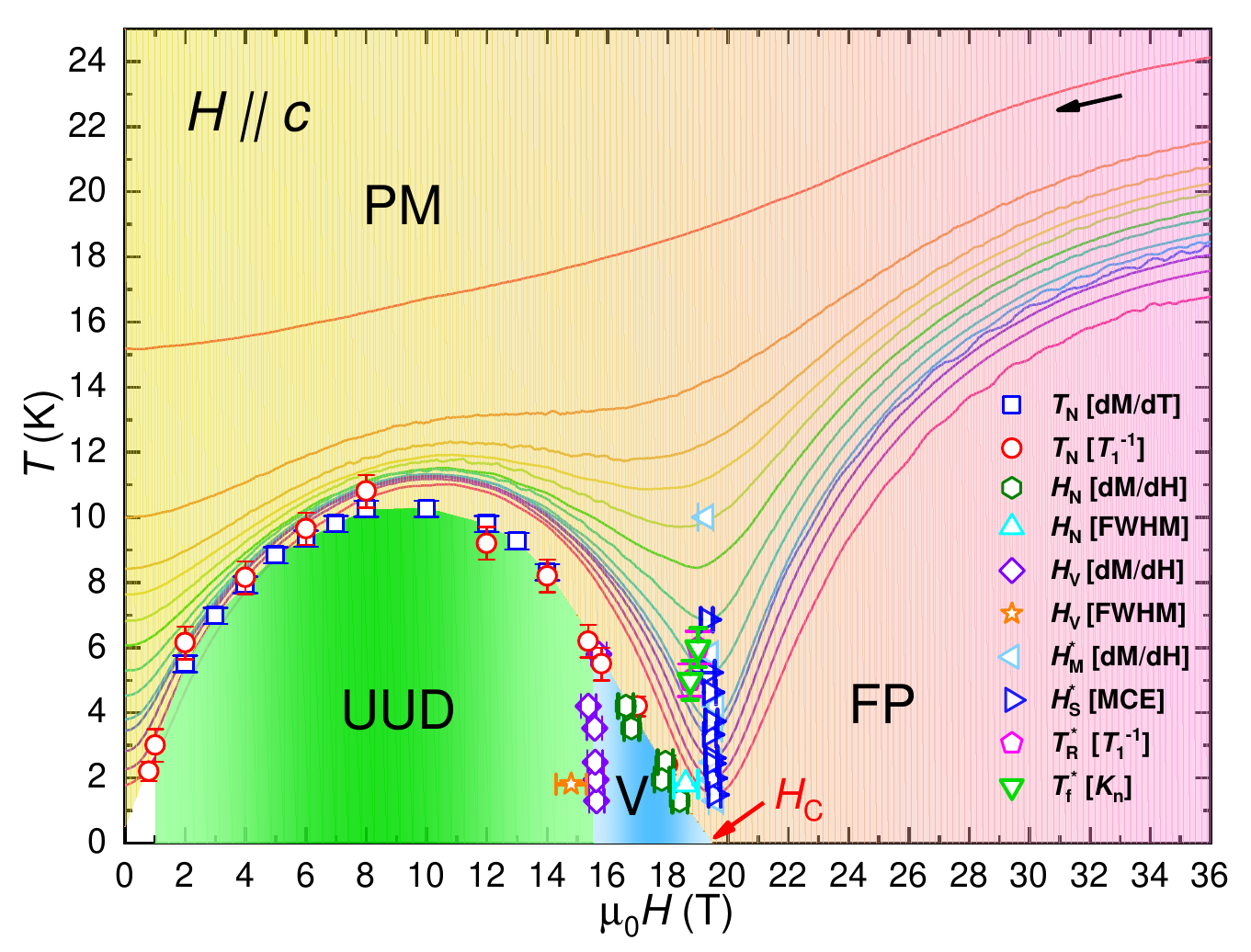}
    \caption{\label{pd}
    {\bf Phase diagram.}
    Solid lines represent the adiabatic $T(H)$ data with field by down sweeps. $H^*_{\rm S}$ denotes the high-temperature dip position in the $T(H)$ curve. Open symbols represent phase transition or crossovers boundaries from different probes, including $T_{\rm N}$ determined in $dM/dT$ and $1/T_1$, $H_{\rm N}$ in $dM/dH$ and FWHM, $H_{\rm V}$ (UUD-V boundary) in $dM/dH$ and FWHM, and high-temperature crossovers $H^*_{\rm M}$ in $dM/dH$, $H^*_{\rm S}$ in adiabatic $T(H)$ data, $T^*_{\rm R}$ in $1/T_1$, and $T^*_{\rm f}$ in $K_{\rm n}$.
    }
\end{figure}

In this work, we focus on the high-field phase of Rb$_2$Co(SeO$_3$)$_2$~\cite{ZhongYD_PRM_2020}, an isostructural compound to K$_2$Co(SeO$_3$)$_2$, to investigate the supersolidity and related quantum criticality. By measuring nuclear magnetic resonance (NMR) and pulsed-field magnetization and MCE, we
establish the magnetic phase diagram of Rb$_2$Co(SeO$_3$)$_2$ in fields up to 36~T. The phase diagram consists of a sequence of phases, including a gapped UUD phase, a putative V-type SS
above 15.8~T,  and a fully polarized (FP) phase above $H_{\rm C}~\approx$~19.5~T, as shown in Fig.~\ref{pd}.
Distinct from the UUD phase, the V-type SS features gapless excitations and an elongated temperature range of fluctuations below $T_{\rm N}$.
We further identified a QCP at $H_{\rm C}$ in between the SS and FP phases and revealed a novel critical exponent ${\nu}z~\approx~$1 from the MCE data.
Furthermore, near $H_{\rm C}$, the $M(H)$, NMR, and MCE data provide consistent evidences for persistent spin fluctuations over a broad temperature range. This suggests Rb$_2$Co(SeO$_3$)$_2$ is a promising candidate for magnetocaloric cooling.

{\bf Methods.}
High-quality single crystals of Rb$_2$Co(SeO$_3$)$_2$ were grown by the flux method~\cite{ZhongYD_PRM_2020}.
In this study, the magnetic field was applied along the crystalline $c$-axis, which is the magnetic easy axis.
The DC magnetic susceptibility was measured using a physical property measurement system (PPMS) with temperature down to 1.8~K.
Pulsed-field magnetization measurements were carried out up to 35~T and pulsed-field MCE measurements (under adiabatic conditions)  up to 38~T.
NMR experiments were conducted on $^{85}$Rb nuclei ($I=5/2$, Zeeman factor $\gamma$~=~4.111~MHz/T)
by the spin-echo method with field up to 22~T.
The NMR Knight shift $K_{\rm n}$ is calculated by $K_n=(f/\gamma H-1){\times}100\%$,
where $f$ is the first moment (average frequency) of the center lines of the spectra.
The spin-lattice relaxation rate $1/T_1$ was measured using the spin inversion-recovery method.

\begin{figure}[t]
    \centering
    \includegraphics[width=8.5cm]{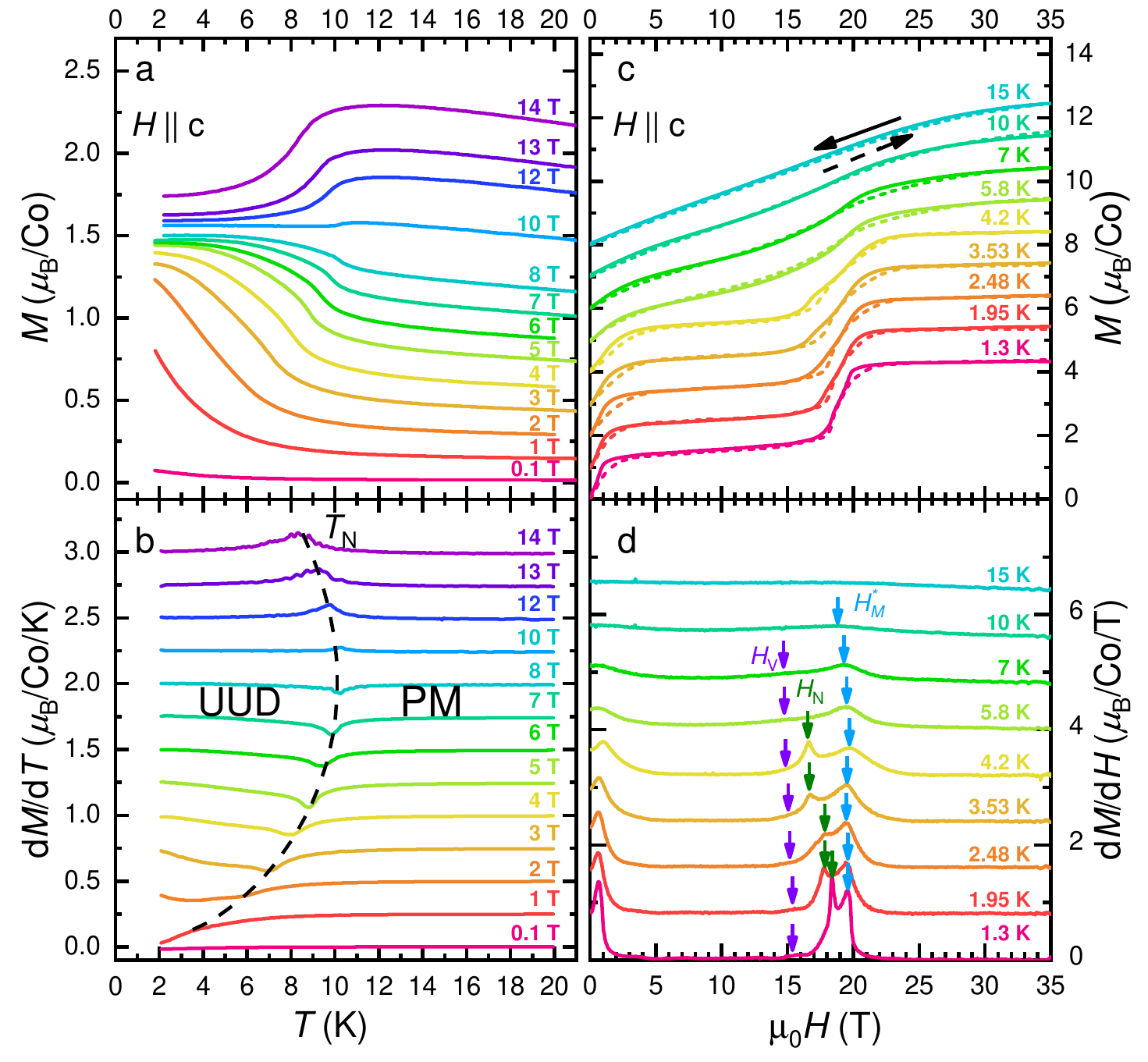}
    \caption{\label{mh}
    {\bf DC and pulsed-field magnetization.}
    {\bf a} $M(T)$ measured under DC field.
    {\bf b} $dM/dT$ as functions of temperatures. $T_{\rm N}$ is determined at the dip and peak position at each field as marked.
    {\bf c} Pulsed-field $M(H)$ measured at selected temperatures with field-up (solid lines) and field-down (dotted lines) sweeps. The 1/3 magnetization plateau is resolved below 4~K.
    {\bf d} $dM/dH$ as functions of fields. All peak features are indicated by down-arrows and marked as $H_{\rm V}$,  $H_{\rm N}$ and $H^*_{\rm M}$, respectively.
    Data are shifted vertically for clarity.
    }
\end{figure}

{\bf DC and pulsed-field magnetization.}
Figure~\ref{mh}{a} presents the DC magnetization $M(T)$ measured with the field from 0.1~T to 14~T. At low temperatures, $M(T)$ exhibits an upturn for $H<10$~T and a downturn for $H>10$~T. By calculating the derivatives of the magnetization, $dM/dT$, as shown in Fig.~\ref{mh}b, the peak and the dip features are seen as connected by the dashed lines. These dip and peak temperatures track exactly the N\'{e}el temperature $T_{\rm N}$,  as confirmed by the NMR spectra shown below.
$T_N$ extracted at each field is then plotted in the phase diagram of Fig.~\ref{pd}. With increasing the field, $T_N$ first increases and then decreases, with
a maximum achieved at about 8~T.

$M(H)$ curves in pulsed magnetic fields are presented in Fig.~\ref{mh}c, with temperature from 1.3~K to 15~K.
A weak hysteresis is observed among field-up and field-down sweeps, which is probably non-intrinsic and is attributed to
the difficulty of achieving full thermal equilibrium between the sample and the bath in pulsed-field~\cite{Zavareh_PRA_2017}.
At 1.3~K, a clear 1/3 magnetization plateau is observed between 2~T and 15.8~T, characteristic of an UUD phase~\cite{ZhongYD_PRM_2020}.
Above 20~T, the FP phase is reached, as indicated by the saturation of the magnetization
to approximately 4.4~$\mu_{\rm B}$/Co$^{2+}$. This gives a large $g$-factor $g\approx$~8.8 due to strong spin-orbit coupling.

To identify all transitions and crossovers at high fields, the differential magnetization $dM/dH$ was derived from the down-sweep curves, as shown in Fig.~\ref{mh}{d}.
At each temperature, distinct peaks are resolved as marked by arrows, labeled with $H_{\rm V}$, $H_{\rm N}$ and $H^*_{\rm M}$, respectively. These peak positions, summarized in the phase diagram of Fig.~\ref{pd}, correspond to phase transitions or crossover fields, as further confirmed by our NMR measurements shown below.

$H_{\rm N}$ and $H_{\rm V}$ are attributed to the N\'{e}el transition and
the left boundary of the V-type SS phase, respectively.
A QCP at $H_{\rm C}~{\approx}$~19.5~T is then determined by extrapolating
$H_{\rm N}$ to zero temperature.  $H_{\rm V}$ varies little with increasing temperature, as indicated by the nearly vertical line in
Fig.~\ref{pd}, until it becomes indistinguishable with $H_{\rm N}$ around 7~K. Interestingly, $H^*_{\rm M}$ can also be extrapolated to $H_{\rm C}$ in the limit of zero temperature, and the prominent peak at $H^*_{\rm M}$ persists up to around 10~K, well above the N\'{e}el transition (see Fig.~\ref{pd}). This suggests $H^*_{\rm M}$
to be a crossover line characterizing critical fluctuations near the QCP at $H_{\rm C}$, as will be confirmed by our MCE data shown below. Notably, similar crossover was also
reported in K$_2$Co(SeO$_3$)$_2$~\cite{ChenT_arxiv_2024}.

\begin{figure*}[t]
    \centering
    \includegraphics[width=17cm]{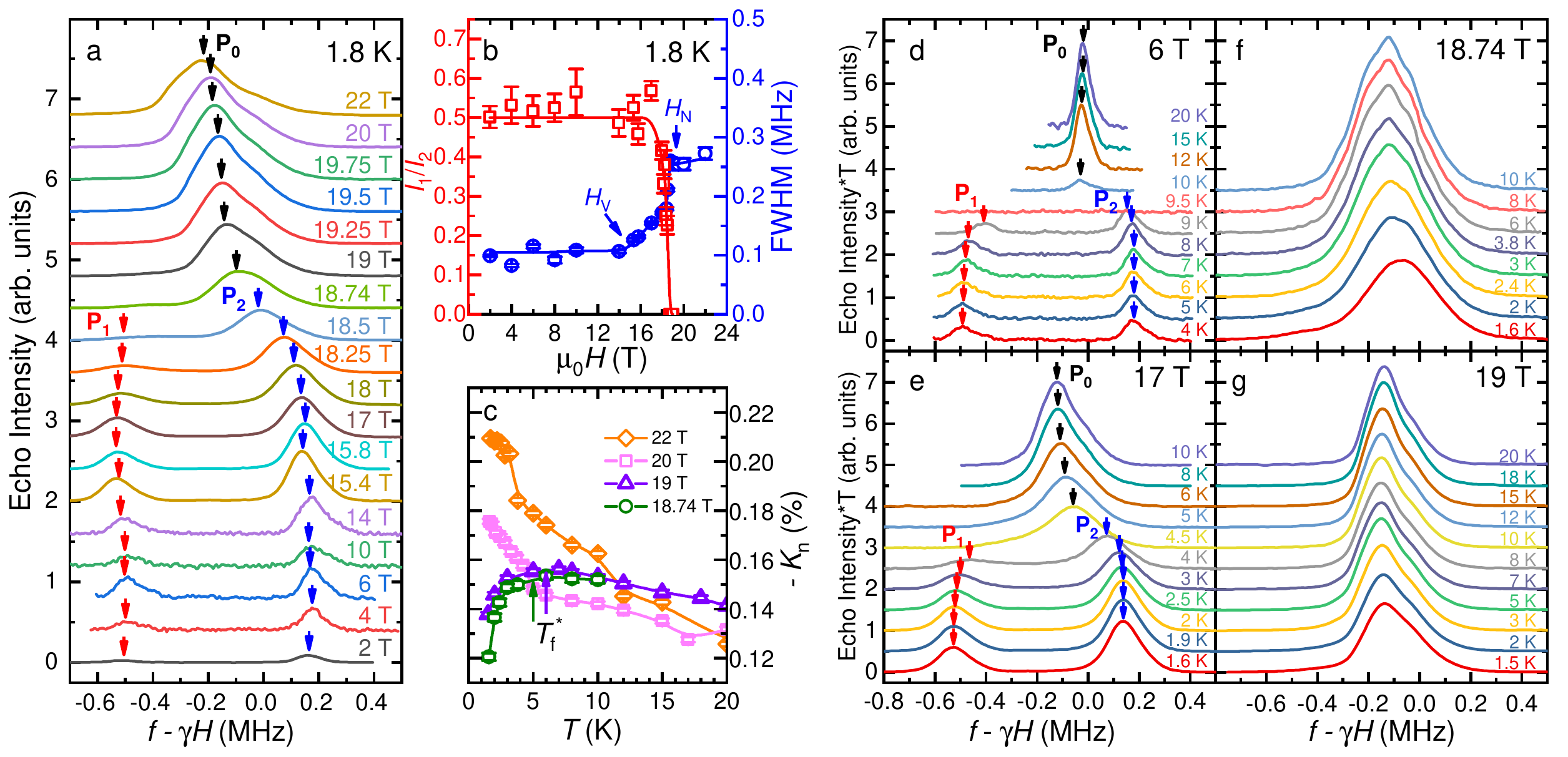}
    \caption{\label{spec}
    {\bf $^{85}$Rb NMR spectra.}
    {\bf a} Center spectral lines measured at 1.8~K under typical fields.
    P$_0$ (above $T_{\rm N}$), P$_1$, and P$_2$ (below $T_{\rm N}$) mark different resonant.
    {\bf b} Relative spectra weight ratio $I_1/I_2$ (left axis) and FWHM of P$_0$  and P$_2$ (right axis) as a functions of field taken at 1.8~K.
    {\bf c} $K_{\rm n}(T)$ at fields close to $H_{\rm C}$.
    {\bf d-g} Center spectral lines measured at typical fields and temperatures, which characterize the UUD phase in {\bf d}, the SS phase in {\bf e}, and the critical regime in {\bf f} and {\bf g} at low temperatures. Spectra are shifted vertically for clarity.
        }
\end{figure*}

{\bf NMR spectra.}
Figure~\ref{spec}{a} shows the NMR spectra at 1.8~K under different fields. In the FP phase, a single resonance peak, denoted P$_{\rm 0}$, is observed  as shown at $H\approx$22~T, indicating uniform magnetization.  At fields below 18.74~T, the spectrum splits into two distinct peaks, labeled as P$_1$ and P$_2$ respectively.  This splitting characterizes the onset of AFM ordering,
as it generates two inequivalent hyperfine fields at the $^{85}$Rb nuclei sites located above the Co$^{2+}$ ions.
The relative spectral weight between the two peaks, $I_1/I_2$, is calculated and plotted as a function of field
in Fig.~\ref{spec}{b}. Indeed, as the field increases from 2~T to 17~T, $I_1/I_2$ remains constant at 0.5. This value indicates the system is in either the UUD or V phase, in which the distribution of local moments with nearly opposite orientations leads to the 1:2 ratio.

The full width at half maximum (FWHM) obtained from P$_0$ and P$_2$ is also plotted in Fig.~\ref{spec}{b}. The FWHM remains constant at fields up to 15~T.
Above 15~T (denoted $H_{\rm V}$), the FWHM increases dramatically until it saturates
at 19~T (denoted $H_{\rm N}$). The regime with varying
FWHM should determine
the onset of a different magnetic structure intercepting the UUD phase and the FP phase. Although the observed relative spectral weight of P$_{\rm 1}$ is less than 0.5 in this field range (Fig.~\ref{spec}{a}-{b}), we believe this is caused by the finite-temperature fluctuations,
as will be demonstrated later in Fig.~\ref{spec}{e}.

Detailed spectra at typical temperatures with increasing fields are demonstrated in Fig.~\ref{spec}{d}-{g}. At 6~T, a magnetic phase transition marked by the
reduction of spectral intensity is observed at 9.5~K, during which
strong low-energy spin fluctuations wipe out part of the signal. Below the transition
temperature, the NMR line splits into two, signifying the onset of AFM order.
The transition temperatures resolved, $T_{\rm N}\approx$9.5~K at 6~T and $T_{\rm N}\approx$4.5~K at 17~T, are consistent with $T_{\rm N}$ ($H_{\rm N}$) determined from $dM/dT$ ($dM/dH$) (see Fig.~\ref{mh}{b} and Fig.~\ref{mh}{d}). At 17~T, the relative intensity of P$_{\rm 1}$ gradually increases to 0.5 with decreasing temperature to  1.6~K. This behavior suggests
the existence of a strongly fluctuating regime, in contrast to
the sharp transition signature of the UUD phase
observed at 6~T. We will show later that this phase is featured with gapless-like excitations in the $1/T_1$ data.

For fields at 18.74~T and above, no line splitting is observed down to 1.5~K, indicating either the transition takes place at even lower temperature or the system is already in
the FP phase. In this high-field regime, we calculated the temperature dependence of $K_{\rm n}$ and plotted it as a function of temperature in Fig.~\ref{spec}{c}.
Here, the negative $K_{\rm n}$ is plotted to adapt the negative hyperfine coupling constant on the $^{85}$Rb nuclei, as demonstrated by the decrease of the resonance frequency
with field close to the FP phase (see P$_0$ in Fig.~\ref{spec}a). At 20~T and 22~T, a monotonic increase of $-K_{\rm n}$ is a clear signature of the FP phase.
At 18.74~T and 19~T, however, $-K_{\rm n}$ exhibits a broad peak with temperature at about $T^*_{\rm f}\sim$6~K.
This broad peak should signal the onset of short-range AFM order instead of a long-range one with symmetry breaking.

\begin{figure}[t]
    \centering
    \includegraphics[width=8.5cm]{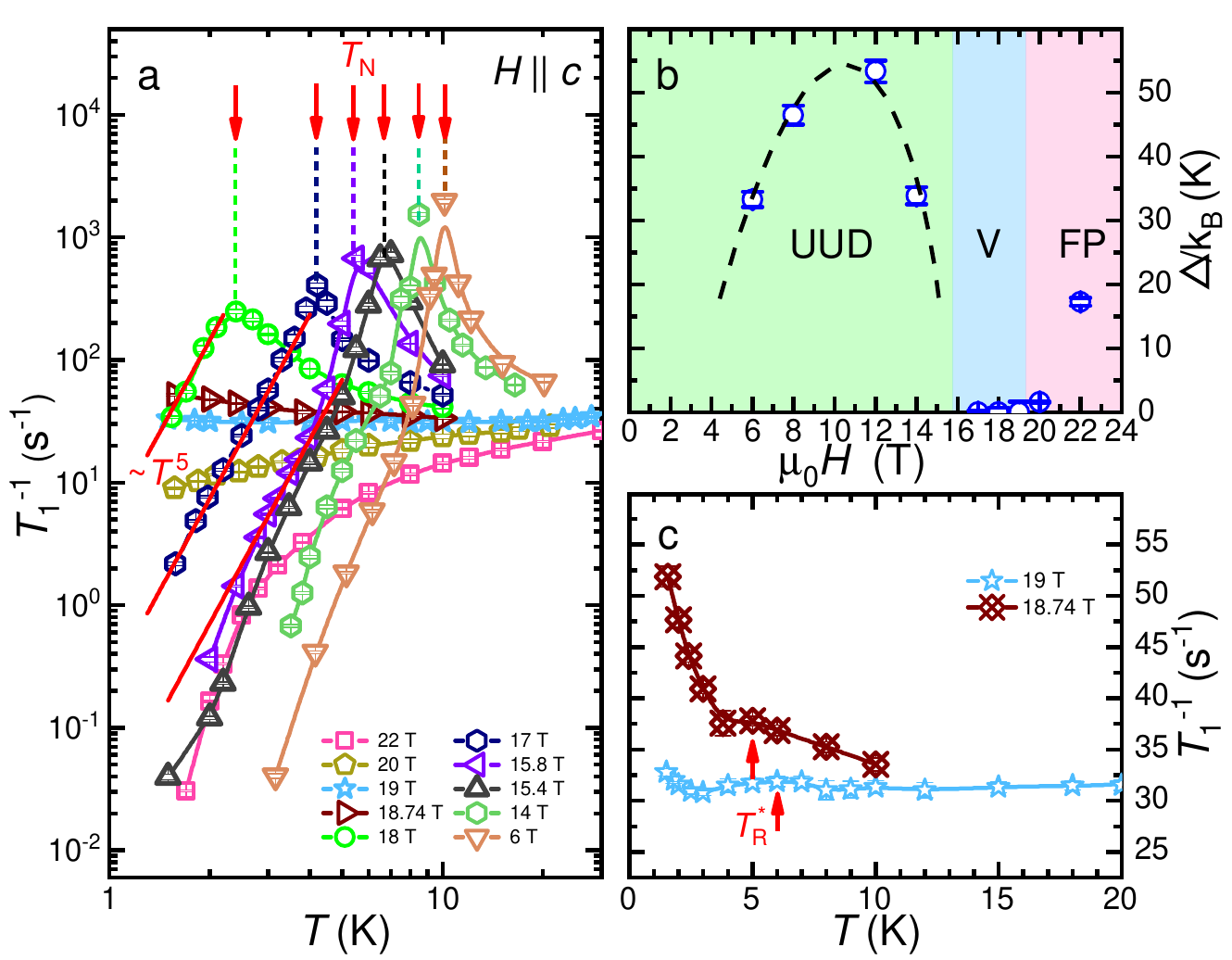}
    \caption{\label{t1}
    {\bf Spin-lattice relaxation rates.}
    {\bf a} $1/T_1$ as a function of temperature, measured under fields from 6~T to 22~T. Peaks marked by red arrows denote AFM transition temperatures $T_{\rm N}$. Solid straight lines
    represent power-law fits $1/T_1 \propto T^5$ at 17~T and 18~T.
    {\bf b} Spin gap $\Delta$ as a function of field extracted by $1/T_1 \propto e^{-\Delta/K_{\rm B}T}$.
    {\bf c} Enlarged view plot of $1/T_1$ at 18.74~T and 19~T. $T_R^*$ mark the temperature location of broad peaks above $T_{\rm N}$.
    }
\end{figure}

{\bf NMR spin-lattice relaxation rate.}
The temperature dependence of $1/T_1$ under several typical fields is
shown in Fig.~\ref{t1}{a}. For fields at and below 18~T, a pronounced peak appears in $1/T_1$ as marked by red arrows, characterizing the magnetic phase transition at $T_N$. As shown in the diagram of Fig.\ref{pd}, $T_N$ values extracted here are consistent with those determined in other measurements.

Remarkably, the low-temperature behavior of $1/T_1$ depends sensitively on the field value. For fields at and below 15~T, $1/T_1$ exhibits a rapid drop below $T_N$, which can be fit to a thermal activation form $1/T_1 \propto e^{-\Delta/k_BT}$ with a gap $\Delta$. This gapped spin excitation suggests that the ground state is a collinear UUD phase. The extracted gap values $\Delta$ are shown in Fig.~\ref{t1}{b}, where $\Delta$ exhibits a dome-shaped field dependence, reaching a maximum at the midpoint of the 1/3 magnetization plateau (Fig.~\ref{mh}{c}).
For fields above 20~T, a gapped behavior is also observed, arising from the Zeeman effect in the FP phase.

In contrast, in the intermediate field range between 17~T and 18~T,
$1/T_1$ follows a power-law temperature dependence with $1/T_1 \propto T^5$ below $T_N$, as shown
in Fig.~\ref{t1}{a}. This $T^5$ behavior can be attributed to a three-magnon Raman process in a gapless AFM state~\cite{Beeman_PR_1968}, and hence supports the ground state to be a V-typed SS~\cite{Yamamoto_PRL_2014}. However, the behavior of $1/T_1$ at even low temperatures must be examined to obtain conclusive evidence
for gapless excitations and therefore establish the SS phase.

An enlarged view of $1/T_1$ at 18.74~T and 19~T is also presented in Fig.~\ref{t1}{c}. The prominent upturn upon cooling below 3~K in both fields suggests that the system tends to order below 2~K, which deserves verification at ultra-low temperatures.
This also implies that the zero-temperature critical field $H_{\rm C}$ is
above 19~T. In addition, a broad peak near 6~K at 18.74~T and 19~T is identified and marked as $T^*_{\rm R}$, which is consistent with the peak temperature $T^*_f$ in $K_{\rm n}$, again as an evidence of short-range magnetic order above $T_{\rm N}$.

\begin{figure}[t]
    \centering
    \includegraphics[width=8.5cm]{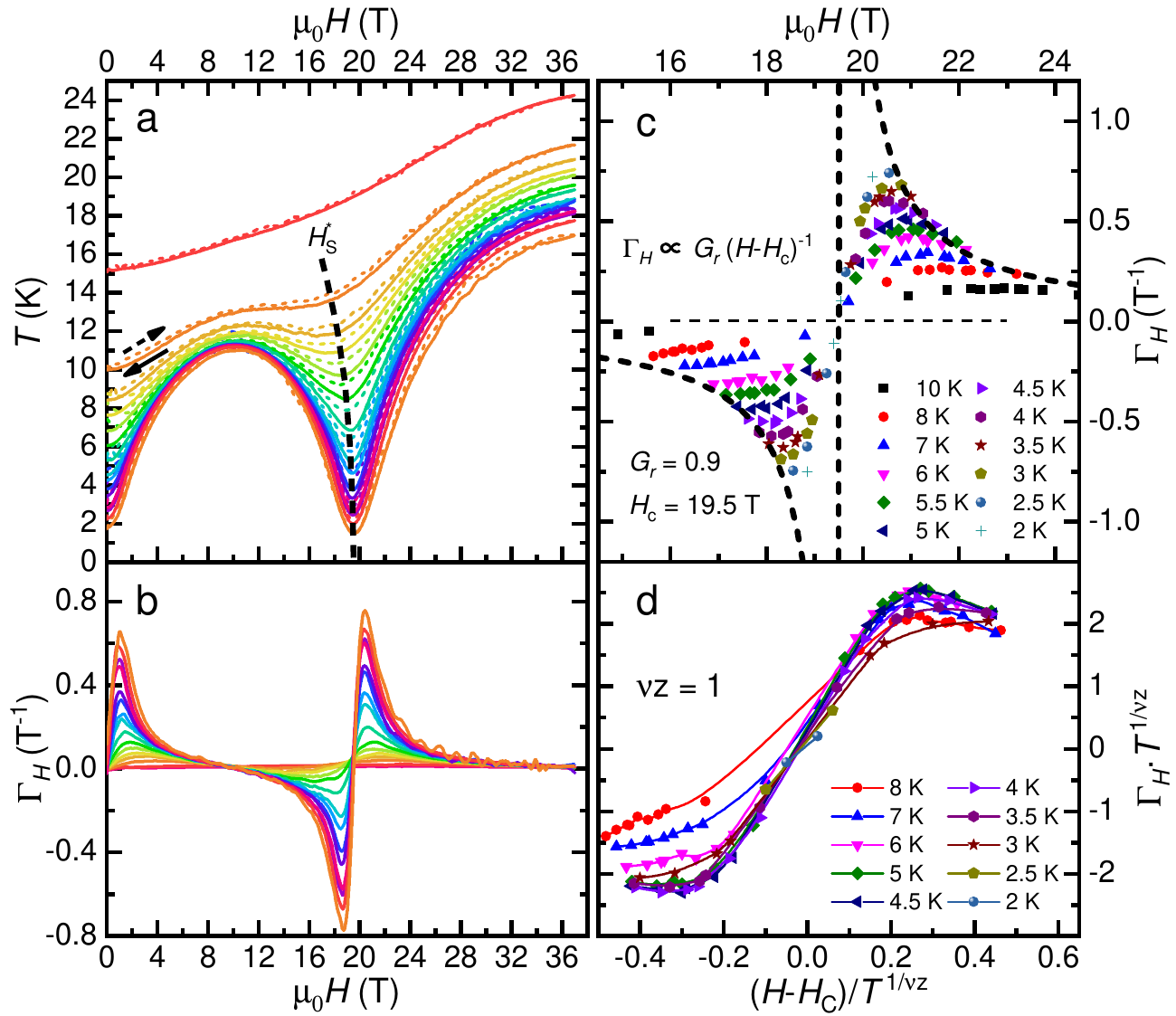}
    \caption{\label{mce}
    {\bf Pulsed-field magnetocaloric data and the Gr\"{u}neisen ratio.}
    {\bf a} Adiabatic $T(H)$ data measured with different initial temperatures, by field-up (dotted lines) and field-down (solid lines) sweeps. $H^*_{\rm S}$ denotes the location of the minimum at each sweep.
    {\bf b} Gr\"{u}neisen ratio $\Gamma_H$ calculated from the $T(H)$ by down sweeps with varying temperatures (see text).
    {\bf c} $\Gamma_H$ plotted as a function of field at selected sample temperatures as listed.
    The dotted line is a function fit $\Gamma_H~=~G_{\rm r}(H-H_{\rm C})^{-1}$ to the envelope of the data, where $H_{\rm C}~\approx~$19.5~T is obtained.
    {\bf d} Data collapse of $\Gamma_H$ in the quantum critical regime which yields  critical exponents ${\nu}z\approx1$.
    }
\end{figure}

{\bf Magnetocaloric effect.}
Pulsed-field MCE measurements were performed under adiabatic conditions~\cite{Kihara_RSI_2013}. The temperature of the sample as
a function of the field under different field sweep directions were recorded and plotted in Fig.~\ref{mce} with different initial temperatures (defined as the sample temperature prior to the field sweep). At
temperature above 10~K or at field above 24~T, the monotonic increase of $T(H)$ with field corresponds to the field polarization effect which causes the entropy release.
With low initial temperatures, two dip features are revealed in $T(H)$ at $H=0$ and $H=H^*_{\rm S}\sim 19.5$~T,
manifesting strong low-energy spin fluctuations at zero field and the critical field $H_{\rm C}$.

In order to study the magnetocaloric cooling effect and quantum critical behaviors between the SS and
the FP phase, the Gr\"{u}neisen ratio, $\Gamma_H=1/TdT(H)/dH$, is calculated and plotted
as a function of field, with selected sample temperatures in Fig.~\ref{mce}b.
As seen in the figure, $\Gamma_H$ exhibits a diverging tendency with decreasing temperature, indicating a large magnetocaloric cooling effect caused by a quantum critical point. In the $T=0$ limit, it can be fit
with $\Gamma_H~=~G_{\rm r}(H-H_{\rm C})^{-1}$ (in Fig.~\ref{mce}c),
with the critical field
$H_{\rm C}~\approx$19.5~T.
In the quantum critical regime, data collapse to a universal scaling function
$\Gamma_H(T)~\sim~T^{-1/{\nu}z} \mathcal{G}[(H-H_{\rm C})/T^{1/\nu z}]$ with the critical exponents ${\nu} z\approx1$
is observed with temperature from 2~K to 6~K, as shown in Fig.~\ref{mce}d.

{\bf Phase diagram and discussions.}
With above magnetization and NMR data, the phase diagram with all phase boundaries and crossovers is constructed and plotted, with
the MCE data overlaid, in Fig.~\ref{pd}.
The $T_{\rm N}$ and the $T_{\rm V}$ values of different probes are determined consistently. For initial temperatures below 4~K, the $T(H)$ curve matches the $T_{\rm N}$ determined by other measurements in the low-field side below 8~T, reflecting dominate entropy release at the transition temperature. The supersolidity with fields from 17 to 18~T is supported by a power-law temperature dependence of $1/T_1$ below $T_{\rm N}$, distinctive from the gapped behavior of the UUD phase (below 15~T).
Further NMR measurements at high fields and ultralow temperatures are necessary to fully settle  the SS phase.

Since the magnetic structures in both the UUD and the V phase contain
local moments aligned along two different directions with a population of 1:2~\cite{Xu_arxiv_2025}, double NMR lines with a relative spectral weight of 1:2 are expected in both phases. However, compared to the UUD phase, the saturation to this 1:2 ratio in
the V phase in Rb$_2$Co(SeO$_3$)$_2$ is
much slower, as shown in Fig.~\ref{spec}{b} and {e}. This is partly due to the weak
interlayer exchange coupling~\cite{ZhongYD_PRM_2020}, which hinders the ordering of the in-plane spin components until ultralow temperature.
Surprisingly, the development of the $z$-component order is also
much slower in the V phase than in the UUD one. This implies either strong
interplay between the in-plane and out-of-plane spin components in the SS, likely via the pseudo-Goldstone mode, or the gap in the UUD phase that protecting the spin order drops substantially approaching the transition to the V phase. Indeed, the slight different $H_{\rm V}$ values determined from the FWHM of NMR spectra and $dM/dH$, along with the little entropy difference between the two phases inferred from the almost vertical $H_{\rm V}$ curve, supports a weakly first-order UUD-to-V-SS transition.

Now we discuss the fluctuation behaviors near the QCP $H_{\rm C}$. The high-field dip
at $H^*_{\rm S}$ of the adiabatic $T(H)$ curve
depicts a crossover line of maximal magnetic entropy arising from the quantum criticality.
The values of  $H^*_{\rm S}$ from MCE match well with
those of $H^*_{\rm M}$ determined in $dM/dH$, and both approach the QCP $H_{\rm C}$ at zero temperature. This suggests that the transition from the SS to the FP phase is through a single QCP where both the in-plane U(1) and the out-of-plane translational symmetries are simultaneously recovered. Note that this is in contrast to the numerical results in 2D models, where a first-order transition was predicted~\cite{Sellmann_PRB_2015,Yamamoto_PRL_2014}.
In our 3D system, the scaling of $\Gamma_H$ gives $\nu z\approx1$, which is consistent with $\nu=1/2$ and $z=2$. This implies that the effective dimension $d+z=5$ is beyond the upper critical dimension, and hence the transition is a continuous one which is well described by the mean-field theory.

However, strong spin fluctuations persist above some characteristic temperature as observed in NMR measurements:
$T^*_{\rm f}$ in $K_{\rm n}$ and $T^*_{\rm R}$ in $1/T_1$ both indicate the onset of a short-range order
above $T_{\rm N}$. This is a phenomenon commonly seen in quasi-2D
frustrated magnets, which is driven by the significant release of entropy
from competing phases at finite temperatures ~\cite{Cui_Science_2023,Cui_CPL_2025}. Note that the fields where $T^*_{\rm f}$ is observed is close to $H^*_{\rm S}$, suggesting spin fluctuations in this regime are influenced by both quantum criticality and magnetic frustration.

This interplay gives rise to a large MCE near $H_{\rm C}$:
As shown in Fig.~\ref{mce}a, the lowest line of the adiabatic $T(H)$ approaches an even lower value at $H_{\rm C}$ than that at zero-field.
This makes Rb$_2$Co(SeO$_3$)$_2$ a promising candidate platform for cryogenic magnetocaloric cooling at high fields.

{\bf Summary.}
In this work, the Ising triangular-lattice antiferromagnet Rb$_2$Co(SeO$_3$)$_2$ is investigated with high-field magnetization, MCE, and NMR measurements. The resulting $H$-$T$ phase diagram contains a gapped UUD phase at low fields and a V-type spin supersolid
with gapless excitation.
The gradual development of the spin order in the V-type SS
reveals a very weak interlayer exchange coupling and a possible interplay of the in-plane and the out-of-plane spin components. We identified a single QCP in the (3+2)D mean-field universality in between the SS and the FP phase.
Near the QCP, there is a finite-temperature crossover regime with strong magnetic fluctuations originating from
a combined effect of magnetic frustration and quantum criticality. The significant entropy accumulation in the melting of the spin supersolid state near the QCP highlights the potential of this compound in application for magnetocaloric cooling.

Note added: During preparation of the manuscript, we are aware of a recent
work on the same compound~\cite{MaLong_arxiv_2025}.

{\bf Acknowledgments.}--- This work is supported by the National Key Research and Development Program of China (Grant No. 2023YFA1406500), the Scientific Research Innovation Capability Support Project for Young Faculty (Grant No.~ZYGXQNJSKYCXNLZCXM-M26), and the National Natural Science Foundation of China (Grant Nos.~12374156, 12134020, and 12334008). The pulsed-field measurements were taken at the high-field laboratory of the Institute for Solid State Physics (ISSP), University of Tokyo. The high-field NMR measurements were taken at
Synergetic Extreme Condition User Facility (SECUF), https://cstr.cn/31123.02.SECUF.


%

\end{document}